\documentclass[showpacs,preprintnumbers,amsmath,amssymb]{revtex4}
\usepackage{graphicx}
\usepackage{dcolumn}
\usepackage{bm}
\begin{document}
\title{Experimental Study on the Conditions for Observing the Josephson Effect as a Method for Finding the Right Theory of Superconductivity}
\author{I.M.Yurin}
 \email{yurinoffice@mail.ru}
\affiliation{%
I.M.Yurin, Fl.61, bld. 7, 22 Festivalnaya St, Moscow, 125581, Russian Federation}%

\date{\today}

\begin{abstract}
The article suggests an experiment enabling to find the right theory
of superconductivity.
\end{abstract}

\maketitle

\section{\label{sec:level1}Introduction}
The Josephson effect (JE) holds a special place in theoretical
physics. It could hardly be denied that it was the prediction
\cite{1} of this effect, as well as its subsequent interpretation
\cite{2} and observation \cite{3,4}, that led to the recognition of
the BCS theory \cite{5} as the standard theory of superconductivity.

However, in 1964, doubts arose in the scientific community about the
consistency of the JE interpretation in the BCS theory. The essence
of these doubts was as follows.

If parameters $\left( {u,v} \right)$ in the Bogoliubov
transformation are replaced with $\left( {u\exp (i\varphi ),v\exp (
- i\varphi )} \right)$, the equations for them within the limits of
the variational procedure existing in the BCS theory will still be
valid. This is not surprising, because such a replacement
corresponds to the unitary transformation $\exp (i\varphi \hat N)$
of the basis set of the system wave functions, where $\hat N$ is the
operator of total number of electrons in the system.

On the other hand, consider the two superconductors \cite{2}
contacting through the tunneling Hamiltonian
\begin{equation}
\hat H_T  = \sum\limits_{\sigma ,\mathbf{k},\mathbf{q}}
{T_{\mathbf{kq}} c_{\sigma \mathbf{k}}^ +  c_{\sigma \mathbf{q}} } +
H.c.
\label{eq:Ht}
\end{equation}
where operators corresponding to the momentum $\mathbf{k}$ are
related to the electrons of the first superconductor, and those
corresponding to the momentum $\mathbf{q}$ are related to the
electrons of the second superconductor.

If we apply transformation $\exp \left( {i\varphi _1 \hat N_1 }
\right)$ to the basis set of wave functions of the first
superconductor, and transformation $\exp \left( {i\varphi _2 \hat
N_2 } \right)$ - to that of the second superconductor, then it is
possible to calculate the second-order correction to the energy of
the system of two superconductors, which appears to depend on the
phase difference $\varphi _2  - \varphi _1$. This is contrary to the
principles of quantum mechanics, because the values of observables
should not depend on the choice of the basis set of the wave
functions of systems \cite{6}.

The analysis shows that the cause of the described paradox is the
breaking of the U(1) symmetry of the original Hamiltonian upon the
transition to the single-particle Hamiltonian written in the
creation and annihilation operators of so-called bogolons.
Unfortunately, proponents of the BCS theory decided to use the
observed paradox for the interpretation of the JE instead
eliminating the paradox.

Recently, it was found that the theory of He II can be constructed
without using anomalous expectation values \cite{7}. It turned out
that the terms in the expression for Josephson flux resulting from
the breaking of the U(1) symmetry in the Bogoliubov theory \cite{8}
do not appear in the strict theory of He II \cite{7}. The absence of
such terms in the strict He II theory casts doubt both on the
physical status of expressions for Josephson current (JC) in the BCS
theory and on the BCS theory itself. At this point it is appropriate
to note that, apparently, the only superconductivity theory (ScT)
which currently explains consistently the JE \cite{6} has
essentially nothing to do with the basics of the BCS theory. In
particular, the bound states of the electrons in it do not have a
Cooper structure. At the same time the ScT suggested in the series
of works \cite{9,10,11}  has a number of distinct advantages in
contrast to the BCS theory.

Let us start by considering the microscopic base of the BCS theory.
It is assumed in it that there is direct attraction between
electrons due to the exchange of virtual phonons. This assumption
seems naive to the author. This kind of interaction is of the order
of smallness $\sim \sqrt {m/M}$, where $m$ is electron, and $M$ is
ion mass, respectively. In this situation the direct Coulomb
repulsion clearly prevails over such attraction. So, the base of the
BCS theory seems somewhat utopian from the microscopic point of
view.

It is different with the ScT developed by the author. Calculation of
the corrections associated with the influence of the phonon system
upon the electron-electron interaction \cite{9} leads to the
following result for Hamiltonian $\hat H$ of the monoatomic
isotropic metal in a model of nearly free electrons placed in a
positively charged matrix of the ionic subsystem (so that the whole
system is electrically neutral):
\begin{equation}
\hat H = \hat T + \hat H_{ee} ,
\end{equation}
where
\begin{equation}
\hat T = \sum\limits_{\sigma ,{\bf p}} {t_{\bf p} c_{\sigma{\bf p}}^
+  c_{\sigma{\bf p}} } ,
\end{equation}
\begin{equation}
\hat H_{ee}  = \Omega ^{ - 1} \sum\limits_{\sigma{\bf p}}
{\sum\limits_{\nu{\bf k}} {\sum\limits_{\bf q} {U_{\bf q}^{{\bf
p},{\bf k}} c_{\sigma{\bf p} + {\bf q}}^ +  c_{\nu{\bf k} - {\bf
q}}^ +  c_{\nu{\bf k}} c_{\sigma{\bf p}} } } } ,
\end{equation}
\begin{equation}
U_{\bf q}^{{\bf p},{\bf k}}  = {{2\pi e^2 } \over {\varepsilon q^2
}} + \delta U_{\bf q}^{{\bf p},{\bf k}} ,
\label{eq:four}
\end{equation}
\begin{equation}
\delta U_{\bf q}^{{\bf p},{\bf k}}  = {{\pi ze^2 K_F^2 } \over
{3\varepsilon mM}}\left[ {{1 \over {\left( {t_{{\bf p} + {\bf q}}  -
t_{\bf p} } \right)^2  - S_1^2 q^2 }} + {1 \over {\left( {t_{\bf k}
- t_{{\bf k} - {\bf q}} } \right)^2  - S_1^2 q^2 }}} \right],
\label{eq:five}
\end{equation}
$c^+$ and $c$ are creation and annihilation operators, respectively,
of conduction band electrons, $\sigma$ and $\nu$ are electron spin
indexes, $z$ is the number of conductivity electrons per atom,
$t_{\bf p}  = {{p^2 } \over {2m}}$, $S_1$ is a parameter related to
the observed sonic speed of the longitudinal polarization $S$ as
$S_1^2  = S^2  - {{zK_F^2 } \over {6mM}}$, $K_F$ is Fermi vector, $
\Omega=L \times L \times L$, $L$ is the crystal dimension, and
$\varepsilon$ is the statical long-wave limit of the metal
dielectric function caused by interband transitions.
Eqs.~(\ref{eq:four}, \ref{eq:five}) illustrate the above skepticism
with respect to the possibility of the appearance of direct
electron-electron attraction, at least for the simple metals of
periodic table.

In contrast, in the theory developed by the author the attractive
potential really appears. However, this is not the naive direct
attraction of the electrons used in the BCS theory to construct the
Schrieffer function. This potential appears within the eigenvalue
problem for the calculation of the correlation energy of the
electrons at the nodes of the momenta  grid \cite{9}.

Numerical calculations show that the potential of attraction is
quite sufficient for the appearance of bound states \cite{10}. In
the simplest case of the appearance of a single bound state in the
framework of the said eigenvalue problem the Hamiltonian of the
electron system can be simplified to the following form:
\begin{equation}
\begin{gathered}
  \hat H = \sum\limits_{\sigma ,\mathbf{k}} {t_\mathbf{k} C_{\sigma \mathbf{k}}^ +  C_{\sigma \mathbf{k}} }  - \sum\limits_\mathbf{k} {E_b C_{ \uparrow \mathbf{k}}^ +  C_{ \downarrow \mathbf{k}}^ +  C_{ \downarrow \mathbf{k}} C_{ \uparrow \mathbf{k}} }  \hfill \\
   - \frac{1}
{4}\sum {E_b \left( {C_{ \uparrow \mathbf{p}}^ +  C_{ \downarrow \mathbf{k}}^ +   - C_{ \downarrow \mathbf{p}}^ +  C_{ \uparrow \mathbf{k}}^ +  } \right)\left( {C_{ \downarrow \mathbf{k}} C_{ \uparrow \mathbf{p}}  - C_{ \uparrow \mathbf{k}} C_{ \downarrow \mathbf{p}} } \right)}  \hfill \\
  \left\{ {\mathbf{p} \ne \mathbf{k};\left| {\mathbf{p}_\alpha   - \mathbf{k}_\alpha  } \right| \leqslant 2\pi /L,\alpha  = x,y,z} \right\}, \hfill \\
\end{gathered}
\end{equation}
where $E_b$ is the bound state energy, which is in fact the
correlation energy of two electrons at one or adjacent sites of the
momenta grid. It is easy to show that this kind of Hamiltonian leads
to the appearance of a superconducting transition in the system
\cite{11}.

As for the calculation of transition temperature values carried out
in the framework of the ScT \cite{10}, it has unexpectedly revealed
a clear advantage over the BCS theory. Indeed, both theories use
only one adjustable parameter. However, in the BCS theory, this
parameter is the so-called Coulomb pseudopotential \cite{12}, which
is in no way related to any measurable parameter of the system,
i.e., is practically "bare." On the other hand, in the ScT, the
adjustable parameter should be on the order of the average phonon
frequency, and numerical calculation results do confirm this
theoretical prediction \cite{10}. It is quite obvious that, had this
study appeared in the 1960s, the physics of superconductivity could
have an entirely different history.

In such a situation one can hardly believe that the fundamental
problems of the construction of the BCS theory are solved in it. In
addition, the above problems with the interpretation of the JE make
the BCS theory, strangely enough, an obvious outsider in the contest
of two theories.

At the same time one must admit that the ScT has certain problems
too. Thus, if one considers the data on the single-electron
tunneling \cite{13}, one can pay attention to the specific features
of the current-voltage characteristic (CVC) when junction voltage
$V$ satisfies one of the following relations:
\begin{equation}
eV = \Delta _1  + \Delta _2
\end{equation}
or
\begin{equation}
eV = \Delta _1  - \Delta _2 ,
\end{equation}
where $\Delta _1$ and $\Delta _2$ are the superconductivity gaps of
superconductors.

Supporters of the BCS theory associate these features with
singularities of the energy density of states resulting from the
extrema of the spectra of "bogolons". In contrast, the ScT has no
such singularities. So, it would seem that the author has a complex
problem of explaining these features of the CVC of the tunnel
junction.

However, careful consideration of the CVC (see Fig. 4a \cite{14})
leads to the conclusion that the JE significantly affects the
experimental study of the CVC. In fact, in the process of recording
the CVC curve radiation is generated, which is associated with the
AC JE. As a consequence, the junction itself is fed with AC voltage
of frequency $\nu  = 2eV/h$. In such a situation there is a constant
component of the JC, which can distort the CVC pattern corresponding
to the single-particle transitions.

So, before one sets himself the complex task of explaining the
specific features of the CVC of the single-particle transitions in
the ScT, one must first clear the CVC from the influence of AC JE.
For this purpose, let us first point out one more feature of the JE
interpretation in the BCS theory.

Actually, according to this theory JC is possible also for isolated
superconductors coupled only by a tunneling junction. This is
indicated even by the basis of states used when considering the
tunneling Hamiltonian $\hat H_T$ in the framework of the BCS theory
\cite{2}.

In contrast, in the ScT JC for isolated superconductors coupled only
by a tunneling junction is not possible, because U(1) symmetry is
preserved. It is of importance that JC in ScT occurs only when two
sides of the Josephson junction are linked by current states with
energies near the electrochemical potential level of the system,
i.e. the wave function of the system is common for both
superconductors since before introducing tunneling Hamiltonian
\cite{6}. This fact enables choosing experimentally the right
theory. In addition, a specific configuration of including the
tunnel junction in the circuit allows clearing the CVC of the
single-particle transitions from the influence of the AC JE.

\begin{figure}
\includegraphics{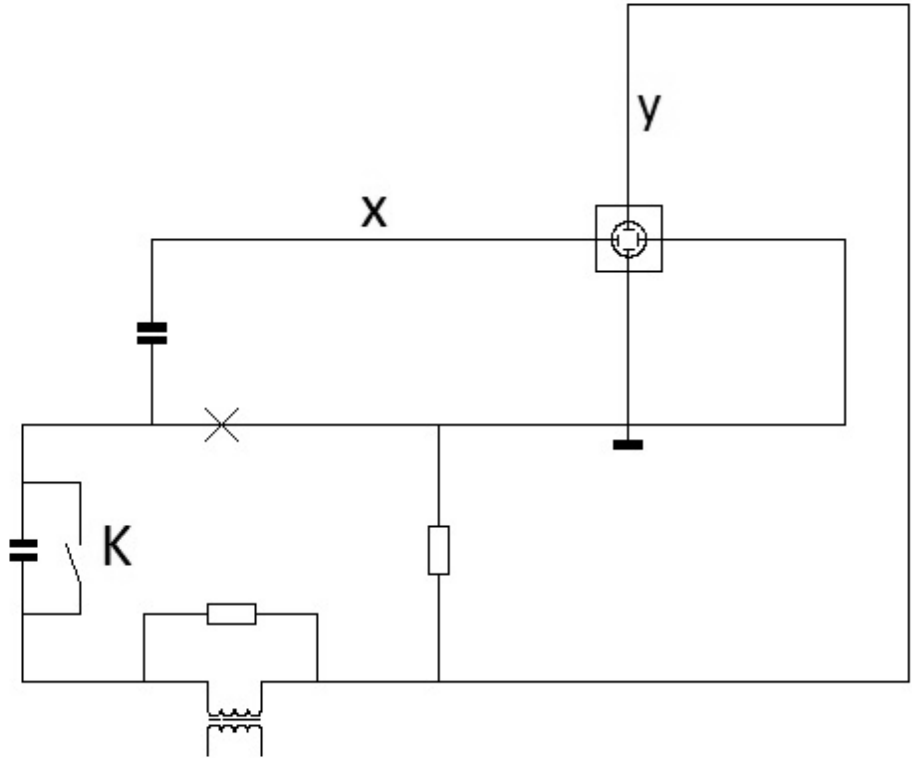}
\caption{\label{fig:epsart}A circuitry of an installation. SIS
tunnel junction is marked with a cross.}
\end{figure}

\section{\label{sec:level1}Experiment Description}
Fig. 1 shows a circuitry of an installation for the study of both
the conditions of the (non)observation of the stationary JE and of
recording the CVC of the single-particle transitions cleared from
the influence of the AC JE.

It is suggested to carry out the measurements with the use of AC. It
can be seen from Fig.1 that, if the switch K is turned off, the left
side of the junction is isolated from the right side by two
capacitors, not taking into account the presence of the tunnel
junction. The signals proportional to the voltage across the tunnel
junction (x) and the current through it (y) are fed to the inputs of
oscilloscope.

If the switch K is turned on, according to the ScT JC is possible.
This means that the CVC will have a region with zero voltage and
non-zero current \cite{3}. In contrast, if the switch K is turned
off, according to the ScT there will be no such region, but the CVC
will only correspond to the single-particle transitions.

\section{\label{sec:level1}Conclusion}
The article suggests an experimental method for finding the right
theory of superconductivity. Recording the CVC of the
single-particle transitions cleared from the influence of the AC JE
will enable to find out to what extent they agree with the
singularities of the energy density of states predicted by the BCS
theory. Carrying out the experiment will also help to answer the
question how much U(1) symmetry breaking is necessary for
constructing the theory of superconductivity. It is important to
note that in the strict theory of He II \cite{7} this breaking looks
obviously optional and in some case lead to errors.

\bibliography{Comparison}

\begin{thebibliography}{}
\bibitem{1} B. D. Josephson, \emph{Phys. Lett.} 1, 251 (1962).
\bibitem{2} P. W. Anderson in \emph{Lectures on the Many-body Problem}, Vol.2, Ed. E. R.
Caianiello (Academic Press, 1964), p. 113.
\bibitem{3} P. W. Anderson and J.M.Rowell, \emph{Phys. Rev. Lett.} 10, 230 (1963).
\bibitem{4} S. Shapiro, \emph{Phys. Rev. Lett.} 11, 80 (1963).
\bibitem{5} Bardeen J., Cooper L.N., Schrieffer J.R. \emph{Phys.Rev.} 106, 162, (1957).
\bibitem{6} I.M.Yurin. Int. Journal of Mod. Phys. B , Vol. 24, No. 30 (2010) 5847-5860.
\bibitem{7} I. M. Yurin, arXiv:1301.5269 (2013).
\bibitem{8} N.N. Bogoliubov. Izv. Akad. Nauk USSR 11 (1947) 77-90.
\bibitem{9} I.M.Yurin, V.B. Kalinin, \textit{Prikl. Fiz. (Rus)}, \textbf{2}, 12
(2003); I.M.Yurin, V.B. Kalinin, arXiv:cond-mat/0102407 (2001).
\bibitem{10} I. M. Yurin, arXiv:cond-mat/0603799 (2006).
\bibitem{11} I. M. Yurin, V. B. Kalinin, arXiv:cond-mat/0202486 (2002).
\bibitem{12} Maksimov E.G., Savrasov D.Yu., Savrasov S.Yu., \emph{Usp Fiz Nauk}, 40, 337 (1997).
\bibitem{13} Giaever I., \emph{Phys.Rev.Letters}, 5, 147, (1960); 5, 464,(1960).
\bibitem{14} Clarke J., \emph{Usp.Fiz.Nauk (USSR)}, \textbf{104} 95-129 (1971).

\end{thebibliography}

\end{document}